# Ag diffusion in SiC high-energy grain boundaries: kinetic Monte Carlo study with first-principle calculations

Hyunseok Ko[a], Jie Deng[b], Izabela Szlufarska[a,b], Dane Morgan[a,b,*]

[a]Materials Science Program, University of Wisconsin-Madison, WI 53706, USA
[b]Materials Science and Engineering Department, University of Wisconsin-Madison, WI 53706, USA

## ABSTRACT

The diffusion of silver (Ag) impurities in high energy grain boundaries (HEGBs) of cubic (3C) silicon carbide (SiC) is studied using an *ab initio* based kinetic Monte Carlo (kMC) model. This study assesses the hypothesis that the HEGB diffusion is responsible for Ag release in Tristructural-Isotropic (TRISO) fuel particles, and provides a specific example to increase understanding of impurity diffusion in highly disordered grain boundaries. The HEGB environment was modeled by an amorphous SiC (a-SiC), which approximately represents the local environments of HEGBs. The structure and stability of Ag defects were calculated using *ab initio* methods based on Density Functional Theory (DFT). The defect energetics suggested that the fastest diffusion takes place via an interstitial mechanism in a-SiC and that is what is modeled. Interstitial sites were identified by a gridded search and transition states were computed using the climbing image nudged elastic band method. The formation energy of Ag interstitials (relative to bulk Ag) and the Kinetic Resolved Activation ($E_{KRA}$) energies between them were well approximated with Gaussian distributions that were then sampled in the kMC. The diffusion of Ag was simulated with the effective medium model using kMC. Ag diffusion coefficients were estimated for the temperature range of 1200-1600°C. Ag in a HEGB is predicted to exhibit an Arrhenius type diffusion and with a diffusion prefactor and effective activation energy of $(2.73 \pm 1.09) \times 10^{-10}$ m$^2$s$^{-1}$ and $2.79 \pm 0.18$ eV, respectively. The comparison between HEGB results to other theoretical studies suggested not only that GB diffusion is predominant over bulk diffusion, but also that the HEGB is one of fastest grain boundary paths for Ag diffusion in SiC. The Ag diffusion coefficient in the HEGB shows a good agreement with ion-implantation measurements, but is 2-3 orders of magnitude lower than the reported diffusion coefficient values extracted from integral release measurements on in- and out-of- pile samples. The discrepancy between GB diffusion and integral release measurements suggests that other contributions are responsible for fast release of Ag in some experiments and we propose that these contributions may arise from radiation enhanced diffusion.

*Corresponding Author. Tel:+1-608-265-5879 ; Email: ddmorgan@wisc.edu (Dane Morgan).

## 1. Introduction

Tristructural-Isotropic (TRISO)-coated fuel is a type of micro-fuel particle that is to be used in the next generation of Very High Temperature Reactors [1]. Individual particles consist of a fuel kernel, typically $UO_2$ but can also be UC or UCO, covered by a C buffer and three coating layers. Considered from the inside out, the coating layers are an inner pyrocarbon layer, a silicon carbide (SiC) layer, and an outer pyrocarbon layer [1]. The SiC layer is designed to provide a structural stability and be a primary barrier for the release of fission products from the fuel to the reactor coolant. The SiC layer is made from the cubic 3C-SiC polytype and has a thickness ~35 µm. Although this layer effectively retains most fission products under the operating and accident conditions, there have been many observations of an undesirable release of metallic fission products under some conditions, particularly radioactive silver ($^{110m}Ag$) [2,3]. Such release of Ag could result in Ag deposition within the reactor and thus raises concerns about the reactor's safety, creates maintenance issues and contributes to restriction on higher operating temperature and associated increased fuel efficiency.

Despite many studies on the topic, the mechanisms for Ag release through the SiC layer of TRISO particle have not been established. Hypotheses for the mechanism of Ag release include chemical degradation of SiC [4,5], vapor diffusion through cracks and nanopores [6-8], and diffusion along grain boundaries (GBs) [9-13]. However, a number of recent studies have provided strong evidence supporting the hypothesis that GB diffusion is a dominant pathway for the silver transport in high-quality CVD-SiC. For example, Friedland *et al.* [8,11] and Gerczak *et al.* [14] recently carried out Ag implantation in a single crystal SiC (sc-SiC) and a polycrystalline CVD-SiC (pc-SiC) at 1300°C. The Ag diffusion was only observed in pc-SiC, but not in the sc-SiC. Lopez-Honorato *et al.* [12,15] designed a diffusion couple model where a layer of silver was trapped between two stoichiometric SiC layers then heat treated up to 1500°C. Significant Ag diffusion into SiC was observed in the samples and transmission electron microscopy images confirmed Ag particles along the columnar GB structure. Most recently, a study by Van Rooyn *et al.* [13] used scanning transmission electron microscopy-energy dispersive X-ray spectroscopy to observe Ag in both GBs and the triple junctions from irradiated TRISO coated particles.

The above considerations suggest that GB diffusion is a preferred mechanism for Ag transport, yet it remains unclear which GBs transport the Ag and how quickly, how this transport occurs at the atomistic level, and if there is any coupling to irradiation effects. To better elucidate the fundamentals of Ag diffusion in SiC, atomistic simulation studies have modeled Ag diffusivity through bulk [16], $\sum$3-GBs [17], and $\sum$5-GBs [18] of 3C-SiC. The diffusion coefficients ($D$) were predicted to be $D_{Bulk}$= $3.9\times10^{-29}$ $m^2s^{-1}$ at 1600°C in bulk SiC (the fastest mechanism being Ag interstitials, where the Ag residues on a site tetragonally coordinated by four Carbon atoms), $D_{\sum3\text{-}GB} = 3.7\times10^{-18}$ $m^2s^{-1}$ at 1600°C in $\sum$3-GBs (for the fastest direction, which was along the $[0\bar{1}1]$), and $D_{\sum5\text{-}GB} = 0.22 - 10.5\times10^{-18}$ $m^2s^{-1}$ at 1227°C in the $\sum$5 (120) tilt GB. These results strongly suggest that the bulk diffusion cannot account for the experimentally observed release rates of Ag from TRISO particles, which are summarized in Table 2. Additionally, the higher $D$ values found for GBs provide further evidence that GB diffusion is a dominant mechanism responsible for Ag release.

Despite the higher $D$ values predicted for select GBs compared to bulk, the predicted $D$ in $\sum$3 and $\sum$5 GBs are still from one to three orders of magnitude lower than the lowest values measured from integral release measurement [9,19-24] at similar temperatures, and therefore they cannot be simply invoked as the explanation for the observed diffusion. A major missing part of present understanding is that high-energy GBs (HEGBs) have not yet been modeled or measured explicitly, but are expected to play a significant role in Ag transportation in pc-SiC. The HEGBs are often highly disordered structures and represent > 40% of GBs in TRISO prototype materials [14,25]. The high fraction of HEGBs allow them to provide a percolating path for Ag transportation, and HEGBs are the one of few GB types that are present in high enough concentration to form a percolating path [17]. More importantly, it is often found that disordered (amorphous) materials provide a faster transportation pathway for extrinsic defects [26-28] compared to crystalline materials. Therefore, we expect HEGBs are the most likely GB type to dominate Ag transport, and the Ag diffusion coefficient in HEGBs to be faster than those in other GBs or bulk, which possibly bridges the remaining discrepancies with integral release measurements. In this work, we used an *ab initio* based stochastic modeling approach to predict the Ag diffusion coefficient in HEGBs.

It is important to note that the Ag diffusion through a pc-SiC may depend not only on the diffusivity in a given GB but also on the microstructure topology, particularly how the GB types are connected to form a network. In this work we focus on estimating $D$ in HEGBs, a specific type of GB, and no model of mesoscale diffusion through a GB network is considered (please see Refs [29,30] for mesoscale transport models). However, since the HEGBs form a percolating network, if they are found to a fast diffusion path we expect these mesoscale effects on the overall measured diffusion coefficient and its activation energy to be relatively minor [29]. Therefore, in this work we will make direct comparisons between our calculated $D$ values for the HEGB and experimentally measured $D$ values from pc-SiC samples.

## 2. Methods

### 2.1. Structure

The HEGB is modeled as an amorphous SiC (a-SiC) region, as the local environments in HEGBs of covalent materials are known to be similar to amorphous phases [31,32] and a bulk amorphous phase is computationally more tractable to model than a full GB structure. We prepared an a-SiC structure from classical molecular dynamics with the Tersoff interatomic potential using the melt-quench method [33]. A supercell containing 128 atoms with a perfect stoichiometry (but allowing antisite and coordination defects) was arbitrary cut out from the bulk, then fully relaxed using Density Functional Theory (DFT) with periodic boundary conditions and under constant zero pressure and zero K temperature. The resulting cell vectors were [11.21, 0.17, -0.37; 0.17, 11.57, 0.15; -0.38, -0.15, 11.03] (non-cubic) and the final density of the cell was 2.98 g/cm$^3$, which is comparable with the simulated density by Tersoff potential (3.057 and 2.896 g/cm$^3$) in Ref [33], and is, as expected, lower than the DFT calculated sc-SiC density, 3.18 g/cm$^3$. It is true that some features of real amorphous structures may depend on cooling and processing conditions, but we make the assumption that the present method captures local environments adequately enough to approximate the diffusion rates of Ag through an amorphous-like GB region.

### 2.2. *ab initio* Calculations

The Vienna Ab-Initio Simulation Package (VASP) [34-37], an *ab initio* DFT code, and projector-augmented plane-wave (PAW) method [38,39] were used to relax the final a-SiC and to explore Ag diffusion in the a-SiC structure. The exchange-correlation was treated in the Generalized Gradient Approximation (GGA), as parameterized by Perdew, Burke, and Ernzerhof (PBE) [40]. A single Γ-point *k*-point mesh was used to sample reciprocal space. While this is not highly converged it was necessary to use few *k*-points to enable the large number of required calculations. The Γ-point *k*-point mesh had an error about 15 meV/atom for 128 atom cell compared to 2×2×2 *k*-point mesh, which is expected to be well-converged. Tests on at least five barriers showed that error in energy barrier with respect to *k*-points for a Γ-point vs. 2×2×2 *k*-point mesh is within an acceptable error range of 200 meV/Ag (this corresponds to about a factor of 5× error at 1500K). The energy cut-off was set at 450eV. The convergence for the electron self-consistency cycle was set to $10^{-4}$ eV. For the formation energy ($\Delta E_f$) of defects, we use the following expression[41] : $\Delta E_f = E_{def} - E_{undef} + \Sigma_I \Delta n_I \mu_I$. The $E_{def}$ and $E_{undef}$ are energies of the defected and the undefected cell, $\Delta n_I$ is the change in the number of the atomic species *I* in the defected cell from the number of same species in the undefected cell, and $\mu_I$ is the chemical potential of atomic species *I* relative to its reference energy. The reference states for Si and C are taken as the bulk Si and C VASP energies (which are in turn referenced to the appropriate atomic species energies given as defaults in the pseudopotential files) in their groundstate structures (diamond lattice for Si and graphite for C). These values are $E_{Si}$=-5.44, and $E_C$=-9.20 eV/atom. Throughout this study, Si-rich condition chemical potentials are used for silicon and carbon ($\mu_{Si}$ = -5.44 and $\mu_C$ = -9.65 eV) for consistency. The *ab initio* formation energy of bulk solid phase Ag metal ($\mu_{Ag}$ = -2.82eV) is set to be the chemical potential for Ag. No charged supercells or explicitly charged defects were considered in this study. We believe this approximation is a reasonable as the neutral state for Ag interstitials (which are our focus in this study, as discussed in Sec.3) is stable charge state for Ag in n-type crystalline SiC, as predicted by Shrader, *et al* [16].

To obtain diffusion pathways and the corresponding minimum energy path, the climbing image nudged elastic band (CI-NEB) method was employed. Typically, five images, including initial and final states, were used in the CI-NEB calculations. The initial estimate for the transition state was set as a center image, and additional images are linearly interpolated between Ag site and the center image. The transition states are estimated by the local geometry of the Ag and Si/C atoms. The construction of center images (e.g. centroid of sites consisting of the initial and final Ag sites and all Si and C neighbors) are presented in the Supplementary Material (SM) Sec.A.1.

In order to represent Ag migration barriers independently of the initial and final state energies we use the Kinetically Resolved Activation (KRA) Barrier [42,43] to represent the barrier value. In the KRA the saddle point migration energy ($E_m^{saddle}$) is determined by the energies of the initial and final sites ($E_f^{initial}$ and $E_f^{final}$) and KRA energy ($E_{KRA}$), which is used to as in Eq. (1) and illustrated in Figure 1. The KRA approach provides a compact way to represent the migration energy of a hop in either a forward or backward direction. Furthermore the approach defines $E_{KRA}$ so as to remove the contributions of the initial and final energy to $E_m^{saddle}$, and $E_{KRA}$ is therefore likely to vary less and be easier to represent than $E_m^{saddle}$.

$$E_{KRA} = E_m^{saddle} - (E_f^{initial} + E_f^{final})/2. \qquad (1)$$

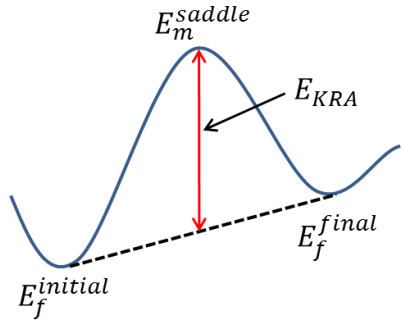

Figure 1. A schematic representation of the Kinetically Resolved Activation (KRA) barrier.

2.3. Kinetic Monte Carlo simulations

In order to explore the diffusion in a disordered lattice we use the effective medium approximation (EMA) [44], which has been applied previously to model diffusion in an amorphous system [45,46]. In this model, a single Ag atom diffuses under the assumption that Ag atoms do not interact with each other while diffusing along HEGB. The main idea of this scheme is to replace the actual system jump frequencies by an effective energy surface with an equal frequency distribution (fitted to DFT calculations), but stochastically sampled. The effective energy surface (effective medium) can be mapped out by creating Ag interstitial sites (lattices), and calculating Ag interstitial formation energies (energy assigned to lattice) and the KRA barriers (transition state energies relative to lattices, which will determine the jump frequency). These parameters are then stochastically sampled until the distributions (μ and σ) in the EMA model are within the standard error from DFT calculations (Table 1).

In the EMA model used in this study, the connectivity between interstitial sites is simulated by adopting the randomly blocked sites approach [47] as depicted in Figure 2 (a). A virtual face centered cubic (*f.c.c.*) lattice was used as an effective medium where each lattice site represents a position of an Ag interstitial in a locally stable state and the number of neighbors on the lattice represent number of nearby minima for Ag interstitial that are connected to the original minimum by a single hop. Then the lattice sites of *f.c.c* are randomly blocked (about 63.33% of sites) until the distribution of remaining nearest-neighbor sites reaches the mean (μ) and standard deviation (σ) of the distribution we obtained from fitting to DFT values. The blocked lattices are marked as inaccessible and not used in the simulation. The hop distance is determined from the distances between Ag interstitials, sampled from DFT, as described in Sec.3.2. Since there was no significant correlation between the migration barriers and Ag interstitial site distances found in the calculations (Sec. 3.1) we assumed that the site energies and the KRA barriers can be chosen independently of site distances. Subsequently each site (e.g., A, B, and C in Figure 2) was assigned a random site energy that is sampled from a fitted distribution of the formation energy (Sec. 3.1). The $E_{KRA}$ (e.g., $E_{KRA}^*$ and $E_{KRA}^{**}$ in Figure 2 (b)) is also sampled from a fitted distribution of DFT barrier values (Sec. 3.1). Now the sampled site and $E_{KRA}$ energies are used to determine the saddle point migration energies (e.g., $E_{A \leftrightarrow B}^{saddle}$ and $E_{B \leftrightarrow C}^{saddle}$) and hop barriers for KMC.

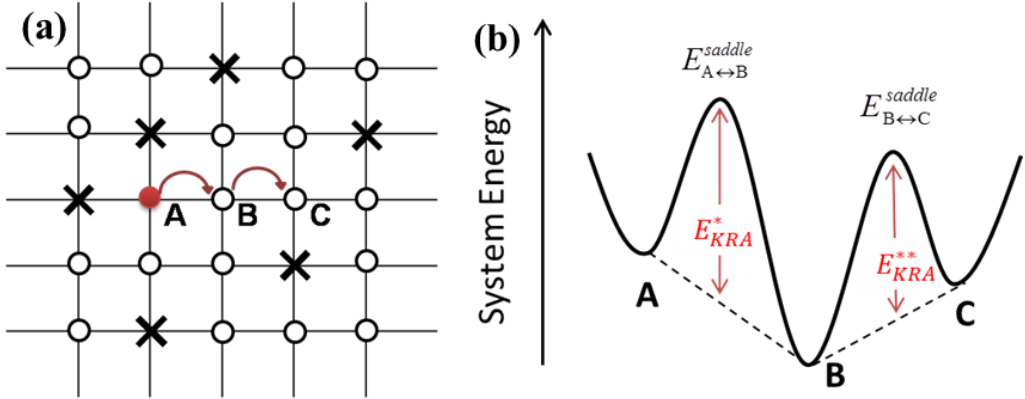

Figure 2. (a) A schematic of the randomly blocked sites model, where cross represents blocked (inaccessible) sites and open circle represents open (accessible) sites. (b) A schematic of the combined scheme of random site energies and random barriers (with KRA) models used to construct the energy landscape when the system evolves from A to B to C. See text for details.

The kinetic Monte Carlo (kMC) method was employed to time evolve the system and observe Ag interstitial diffusion. The simulation was performed on a virtual lattice, using parameters fitted to DFT calculations as described above. Experimentally measured GB dimensions were used for the simulation cell of the virtual lattices (which we will refer to as a virtual HEGB). The virtual HEGB structure should match a typical grain size and HEGB width. With this goal in mind the HEGB structure is set to be a rectangular plate with side of approximately 980nm (~1 μm) and thickness of 1 nm (z-direction) [14,25] in this work. Owing to the low formation energies of Ag defects in HEGB (Sec.3.1.), Ag is assumed to stay within the assumed constant finite thickness of the GBs. Therefore, PBC were applied on the virtual lattice except along the z-direction. The Bortz-Kalos-Liebowitz [48] algorithm was used in the kMC model. The hopping rates ($\Gamma$) for Ag atoms are given by the transition state theory as $\Gamma = \omega \exp(-E_A/k_BT)$, where $\omega$ is attempt frequency and $E_A$ is the activation barrier of the hop. The attempt frequency ($\omega$) can be taken as $\omega = \prod_{3N} v^{equil} / \prod_{3N-1} v^{saddle}$, where $v^{equil}$ and $v^{saddle}$ are the vibrational frequencies at the equilibrium and saddle point, and $N$ is number of atoms in an adequately large region around the diffusion path [49]. We take $N$ to be all atoms within a 3.5 Å cutoff of the Ag in the transition state of the hop being considered, which is typically 12-16 atoms, including the Ag. Tests for longer cutoff distances and larger N gave change in $\omega$ of less than 5%. From *ab initio* calculations for 10 test cases, we have found that the attempt frequencies typically range from $10^{12}$ to $10^{13}$ s$^{-1}$, with an average of $4.5 \times 10^{12}$ s$^{-1}$. As we found no simple correlation between the migration energetics and the attempt frequency (e.g., as has been suggested by the Meyer-Neldel rule applied in 3C-SiC [50]), the $\omega$ value for kMC model is simple taken as our sample average value of $4.5 \times 10^{12}$ s$^{-1}$. The activation barrier ($E_A$) for each hop is estimated by the formalism of $E_{KRA}$ energy associated with energies of two equilibrium sites. In each step of kMC, Ag migrates from the current lattice site ($i$) to one of $Z$ neighbor lattices. The probability for Ag to migrate from lattice site $i$ to $j$ is determined stochastically by $P_{ij} = \Gamma_{ij} / \sum_{n=1}^{Z} \Gamma_{in}$, where $j$ is a positive integer from an interval $(0,Z]$. Next, the system is updated to the $j$ state with an increment of time forward, $\Delta t$, given by $\Delta t = \left(\sum_{n=1}^{Z} \Gamma_{ij}\right)^{-1} \ln(u^{-1})$), where $u$ is a uniform random number between 0 and 1. The Ag diffusion coefficient is determined by the Einstein relation, $D_{Ag} = <r^2_{Ag}(t)> / 2dt$, from the calculated mean square displacement as a function of time [51], where $d$ is the dimensionality of the system in which Ag diffuses (here $d = 2$), $t$ is time, and $<r^2_{Ag}(t)>$ is the mean square displacement of Ag as a function of time. We evaluate the average using the multiple time origin method [52]. For the quasi-two dimension HEGB of interest, $<r^2_{Ag}(t)> = 1/N_t [\sum_{j=1}^{N_t} [[r_x(o_{j+(N/2)}) - r_x(o_j)]^2 + [r_y(o_{j+(N/2)}) - r_y(o_j)]^2]$ where $N_t$ is the number of time origins, and the quantity $r_x(o_{j+(N/2)}) - r_x(o_j)$ is the displacement along the x-direction over the time span between time origin $o_j$ and $o_{j+(N/2)}$ (analogous quantities for the y-direction). The diffusion along the z-direction is negligible within the geometry of HEGB (1nm in thickness with no PBC vs. effectively infinite diffusion in the x-y direction due to PBCs) and therefore is not considered for simplicity. For statistical reliability we choose the number of time origins $N_t$ to be half the total number of time steps, $N/2$. For the given $N$ time-steps, $<r^2_{Ag}(t)>$ is computed for $N/2$ possible time origins between $j$ and $(j + N/2)$ time steps. The kMC simulation is allowed explore the system across multiple periodic boundaries along $x$ and $y$ directions. The converged diffusion coefficient values were extrapolated from a linear fit to a plot of $<r^2_{Ag}(t)>$ vs. $t$. Each simulation was typically performed for $10^{10}$ kMC steps to obtain a well-converged fit.

## 3. Results

### 3.1. Ag interstitials

In order to understand the transportation behavior of Ag in the a-SiC, it is important to determine which Ag defect is stable and will contribute to the diffusion. To determine the dominant Ag diffusion species, first the formation energies of the point defects (vacancies, Ag substitutionals on Si and C lattices, and Ag interstitials) were studied. Our results suggest that vacancy mediated diffusion mechanisms are energetically less preferable compared to the interstitial mechanism, and we therefore consider only the interstitial diffusion mechanism for Ag in the a-SiC. Other considered diffusion mechanisms and related Ag defect energies are discussed in SM Sec.A.2.

To model Ag interstitial diffusion in a-SiC, the energy landscape of Ag interstitials was first investigated. In an amorphous material, locating extrinsic interstitial defect sites is challenging because of lack of long-range order. Many approaches to identify interstitial sites have been proposed in other amorphous systems [53-56]. To achieve a complete list of possible Ag interstitial sites, we have taken a brute force but comprehensive approach to identify possible sites. Specifically, we identified all existing interstitial sites by gridding our entire supercell with a fine uniform grid and relaxing a Ag interstitial at every grid point in the cell. A similar approach was used to study Li diffusion in amorphous oxides [56]. When necessary, gridding points were shifted so that they were at least 1.5 Å from any adjacent host atoms to avoid a numerical instability. The ideal grid density was tested for ¼ of the simulation cell to verify a maximum grid size that could still to capture all existing minima. It was determined that a grid size of 1 Å on a side was adequate to find all the interstitials.

For the a-SiC sample, we investigated $11^3$ grid points and identified 153 interstitial sites, after applying a hierarchical clustering approach for sites close to each other, both energetically and geometrically (details in SM Sec.A.3.). The distribution of formation energies is shown in Figure 3 (a). The formation energies of Ag interstitials in a-SiC had an average value of 3.50 eV and this value is ~7 eV more stable than the most stable neutral Ag interstitial in sc-SiC (10.49eV [16]). The values range from -0.59 to 8.21 eV, with a standard deviation of 1.91 eV. The Ag, which is an impurity with a relatively larger size than those studied in similar amorphous systems [53-55,57-60], turned out to be highly stable as interstitial in a-SiC. In SM Sec.A.4., we further investigated several aspects of relaxed systems in order to understand the stability of Ag in a-SiC. It was observed that the enhanced stability of Ag defects in HEGB vs. bulk is possibly due to large relaxations available to the a-SiC system. Figure 3 (b) shows the Ag distance from one interstitial site to the nearest neighbor site. The figure shows that 95% of interstitial sites have a nearest-neighbor interstitial site within 2.0Å and 100% within 2.75Å.

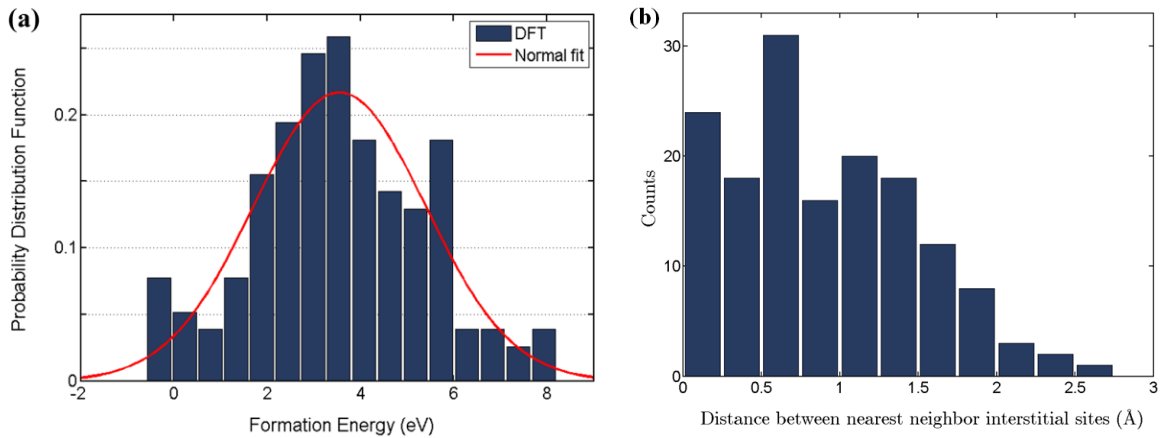

Figure 3. Gridding approach results after clustering (a) Distribution of formation energy with a bin size of 0.5 eV. Comparison to Gaussian distribution parameterized by mean and standard deviation from gridding method. (b) Distribution of Nearest-Neighbor cluster distance from one cluster to the other with a bin size of 0.5 Å.

Migration barriers that are needed for a diffusion analysis of Ag in a-SiC are calculated using the CI-NEB. A number of migration barrier calculations are performed to achieve a statistically robust dataset. Instead of gathering barrier energies for all possible migrations, we focused on elucidating the systematic features of barriers as the sampling size increase. Fifty pairs of minima, approximately 1/6 of the total number of transition states, are randomly sampled. The distribution of migration barriers as a function of the relative Ag distance is shown in Figure 4 (a). The energy barrier for a hop is dependent on the energy state of two endpoints and thus two values (barrier from low energy state to high energy state (L-H) and vice versa (H-L)) are plotted for each reaction. It is observed that there is no correlation between the barrier and the Ag hop distance. However, most of the CI-NEB calculations between sites with distance larger than 2.5Å failed to discover a single saddle point (not shown in Figure 4). They either showed a convergence failure, likely due to significant difference in local ordering that requires movement of multiple atoms, or found two saddle points, which is not a single hop and involves another minimum between them. Therefore 2.5 Å was used as a cutoff distance to determine the number of neighbor interstitial sites (*Z*) from an interstitial site. The *Z* distribution and the normal fit to *Z* are shown in Figure 5. On average, interstitial sites had *Z*=4.4, which represents the average number of sites to which Ag can migrate by a single hop.

The $E_{KRA}$ values as shown in Figure 4 (b) are distributed normally in the range of 0.4-2.2 eV. Table 1 gives the results of fitted normal distributions for Ag interstitial sites for both coordination and energetics. With the KRA formalism and values in Table 1 and Eq.(1) the number of possible hops and their migration energies associated with leaving a given site can be sampled. Therefore, now we can stochastically generate the Ag atom hop energetics that are needed at each kMC step.

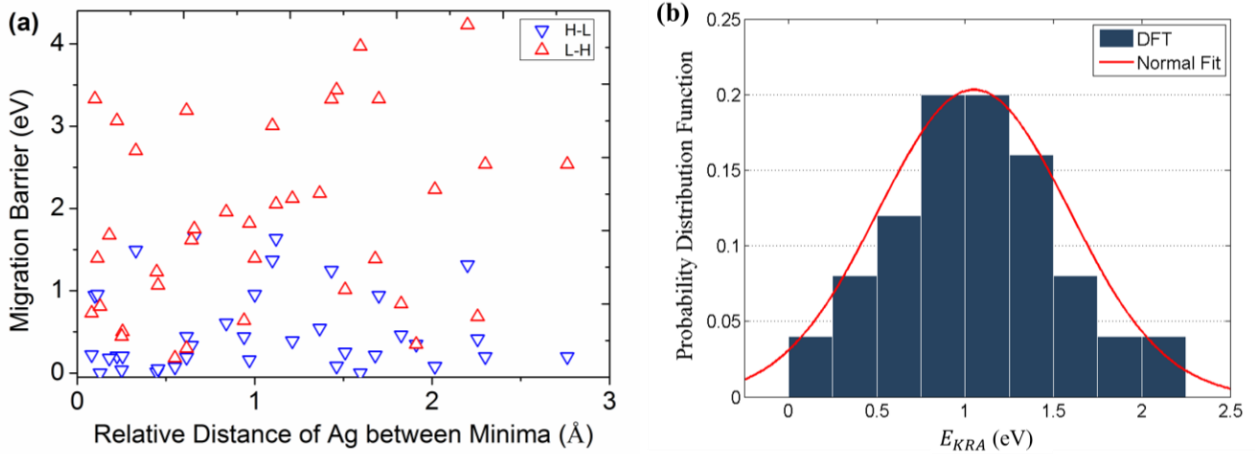

Figure 4. Result of migration barrier calculations for Ag in a-SiC: (a) Energy barriers between two minima plotted as a function of relative distance of minimum. The barrier from high to low and low to high energy state is marked with "H-L" and "L-H" respectively. (b) A histogram of $E_{KRA}$ values from calculated migration barriers in (a).

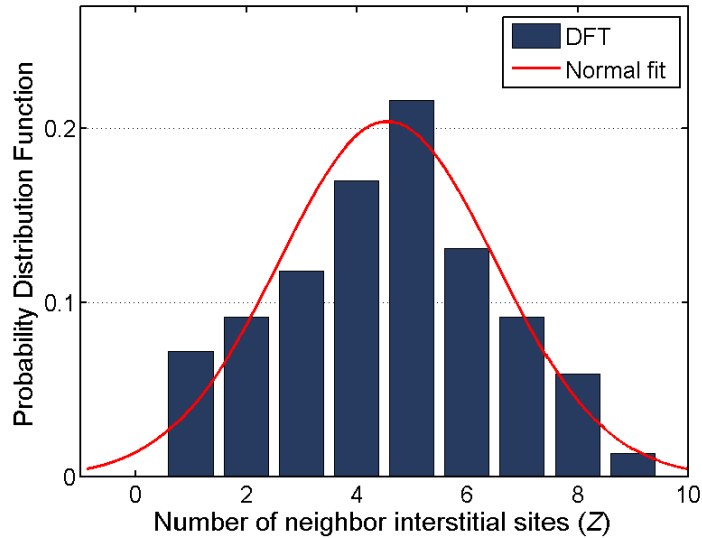

Figure 5. A histogram of the number of neighbor interstitial sites (*Z*) from an interstitial site within the cutoff distance of 2.5 Å of relative Ag positions. A normal distribution fit to the data is shown

with red line.

Table 1. Mean (μ) and Standard deviation (σ) of diffusion related parameters from *ab initio* calculation. All three parameters were fitted to normal distributions (Figure 3 (a), Figure 4 (b), and Figure 5). SE$\mu$ and SE$\sigma$ are the standard errors of mean and standard deviation ((SE$_\mu$ = σ / $\sqrt{n}$, and SE$_\sigma$ ≈ σ/$\sqrt{2(n-1)}$ [61])) from the distribution.

|  | μ | σ | SE$\mu$ | SE$\sigma$ |
|---|---|---|---|---|
| $E_f$ (eV) | 3.50 | 1.91 | 0.15 | 0.11 |
| $E_{KRA}$ (eV) | 1.05 | 0.54 | 0.08 | 0.05 |
| $Z$ (sites) | 4.4 | 2.1 | 0.17 | 0.12 |

### 3.2. Kinetic Monte Carlo model for Ag diffusion

With the statistical distributions from Table 1 we performed a kMC simulation as described in Section 2. To be more specific, we first constructed a virtual medium of *f.c.c.* lattice sites where *f.c.c.* sites correspond to Ag interstitial sites. The single Ag hop distance (lattice distance in EMA) is taken as $\sqrt{\overline{h^2}}$ = 2.1 Å, where $h$ is Ag hop distance sampled from DFT calculation, and $\overline{h^2}$ is the mean of $h^2$ for all sampled Ag hops. Then sites are randomly removed (i.e. become inaccessible) until the distribution of number of remaining nearest neighbor sites agrees with the mean and standard deviation of $Z$, sampled from a-SiC (Table 1). The formation energies are randomly assigned to remaining sites by sampling from the fitted normal distribution of $E_f$ in Table 1. The migration barriers between sites are computed by Eq.(1), where $E_{KRA}$ is sampled from the normal distribution in Table 1. After the Ag interstitial site network has been established, a single Ag is randomly placed on one of the sites and its movement is determined by the rates of migration using kMC. It should be noted that two adjustments to the model are employed in order to model a more realistic effective medium. First consideration was energy correlation between sites. In general we have found a weak correlation between site energy and neighbor energies, but it was found from DFT calculation that low Z sites had higher energies than their neighbors. This correlation is included in EMA. In the calculations without this consideration, the deep-trapping occurred at the low Z sites when their site energies were low. Secondly, the extremum (maximum and minimum) values in sampling $E_f$ and $E_{KRA}$ are limited to the extremum sampled from DFT calculations. In the SM Sec.B.1 details of these considerations are described. The correlations between parameters that were investigated (but not included in the model) are also presented in the SM Sec.B.1.

The typical HEGB dimensions are approximately 1μm×1μm×1nm and we will assume that this is large enough that the Ag diffusivity in a given HEGB is indistinguishable from that which would be obtained from an infinite HEGB, denoted $D_{eff}$. In order to determine $D_{eff}$, we model diffusion in a series of cells each with dimensions of 140nm × 140nm × 1nm, as larger cells were too computationally expensive. To obtain the effective infinite cell diffusivity we calculated the $D_{eff}$ for a number of the 140nm × 140nm × 1nm simulation cells in series. Specifically, $D_{eff}$ is estimated as the harmonic mean of diffusion coefficient in each simulation cell, or $D_{eff} = n / [\Sigma_{i=1}^{n} (D_i)^{-1}]$, where $n$ is the number of sampling ($n=350$) and $D_i$ is the diffusion coefficient from each simulation cell (See SM Sec.B.2. for the standard error of $D_{eff}$).

The calculated $D_{eff}$ in the range from operating to accident temperature of TRISO fuel particles (from 1200 to 1600 °C) are shown in Figure 6 at each temperature. The diffusion prefactors ($D_0$), the effective activation energies ($Q_A$), and the diffusion coefficient ($D$) at 1300 °C and 1600 °C from the literature are also summarized in Table 2 and Figure 6. Ag diffusion simulations were performed in 50 virtual HEGB cells for each temperature. The error bar includes the uncertainty range of $D$ values due to $E_{KRA}$ sampling during kMC (See SM Sec.B.2. for details on the error calculations, where we find that among possible sources of likely error, uncertainty arising from $E_{KRA}$ sampling is the most significant.), and all of the calculated $D$ values are found within the error bars. The diffusion coefficients show Arrhenius type diffusion and the pre-exponential factor ($D_0$) and effective activation barrier ($E_{A,eff}$) are extrapolated from the Arrhenius relation $D_{eff}^{total} = D_0 \cdot \exp(-E_{A,eff} / k_B T)$, where $k_B$ is the Boltzmann constant. The effective activation barrier ($E_{A,eff}$) and diffusion prefactor ($D_0$) predicted by our model are $2.79 \pm 0.18$ eV and $(5.78 \pm 1.09) \times 10^{-10}$ m$^2$s$^{-1}$, respectively. This fitted Arrhenius relation is shown in Figure 6 and shows an excellent agreement with the calculated values. In Figure 7, the effective activation barrier is compared with reported experimental values (open symbol=integral release, and filled symbol = ion implantation), and theoretical predictions (filled triangles) in different structures of SiC.

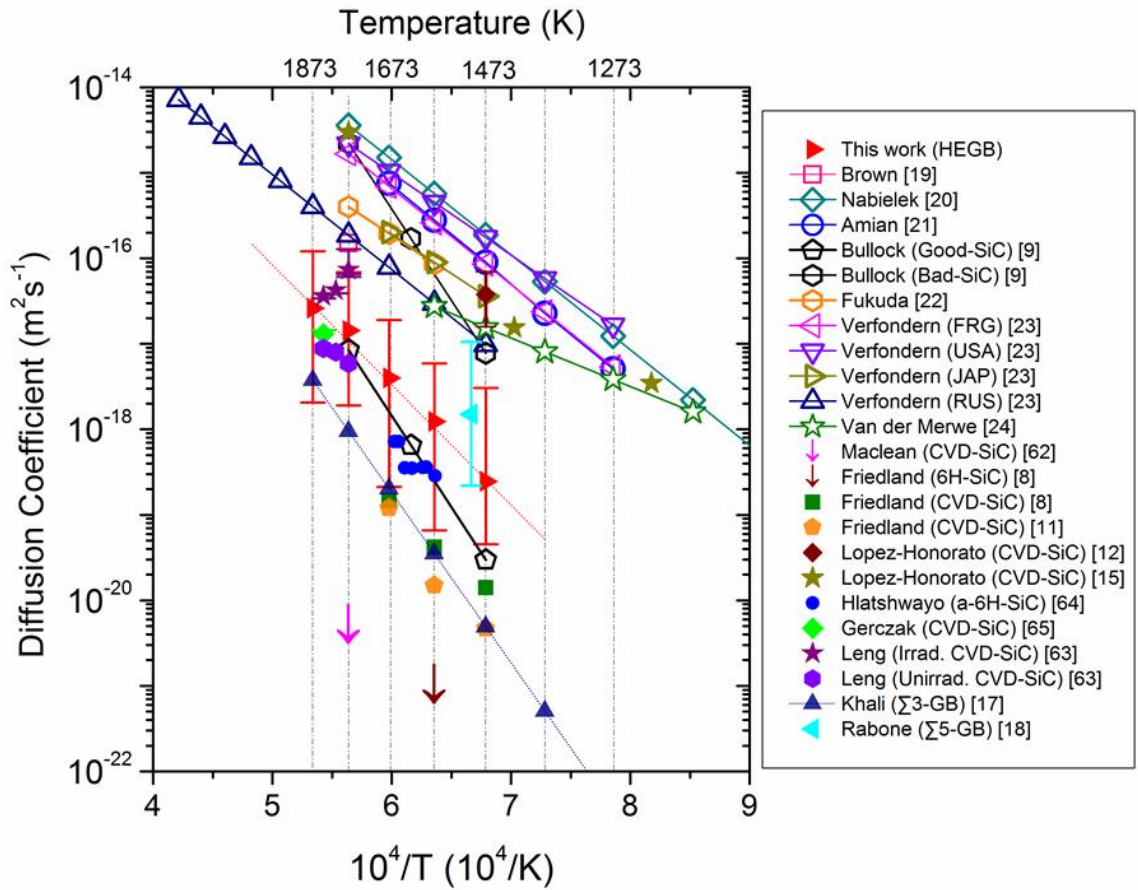

Figure 6. The temperature dependence of Ag diffusion coefficients. Ref [17,18] are the upper bounds for $D$ in crystalline 3C-SiC and $\sum$3-GB from computational study. Ref [9,19-24] are integral release data from irradiated TRISO particles. Open symbols are Arrhenius fit of measurements from irradiated TRISO particles and filled symbols represents reported values from non-irradiated SiC from both surrogate experimental and computational studies. Note the downward arrows are indications that the value is upper limit (i.e. the estimation of $D$ is less than this value). The Arrhenius fit for Ag diffusion in a HEGB from this work is shown with red dashed line.

Table 2. A summary of reported diffusion coefficients for Ag in SiC in the form of an Arrhenius relation $D = D_0 \times \exp(-Q_A / k_B T)$ when available. The temperature range shown gives the range of temperature values used to fit the Arrhenius relation for $D$.

| Method and Reference | Temp.(°C) | $D_0 (m^2 s^{-1})$ | $Q_A$ (eV) | $D$ (T=1300°C) | $D$(T=1600°C) |
|---|---|---|---|---|---|
| **Experiments** | | | | | |
| *Integral Release* | | | | | |
| Brown [19] | [a]1500 | - | - | - | [a]$1.5 \times 10^{-16}$ |
| Nabielek[20] | 800-1500 | $6.8 \times 10^{-9}$ | 2.21 | $5.7 \times 10^{-16}$ | - |
| Amian [21] | 1000-1500 | $4.5 \times 10^{-9}$ | 2.26 | $2.8 \times 10^{-16}$ | - |
| Bullock [9] | 1200-1500 | $9.6 \times 10^{-6}$ | 4.24 | - | - |
| Fukuda [22] | 1300-1500 | $6.8 \times 10^{-11}$ | 1.84 | - | - |
| Verfondern (FRG)[23] | 1000-1500 | $3.6 \times 10^{-9}$ | 2.23 | $2.6 \times 10^{-16}$ | - |
| Verfondern (USA) [23] | 1000-1500 | $5.0 \times 10^{-10}$ | 1.89 | $4.5 \times 10^{-16}$ | - |
| Verfondern (JAP) [23] | 1200-1400 | $6.8 \times 10^{-11}$ | 1.84 | $9.0 \times 10^{-17}$ | - |
| Verfondern (RUS) [23] | 1200-2300 | $3.5 \times 10^{-10}$ | 2.21 | $3.0 \times 10^{-17}$ | $4.0 \times 10^{-16}$ |
| Van der Merwe [24] | 920-1290 | $1.14 \times 10^{-13}$ | 1.13 | $2.7 \times 10^{-17}$ | - |
| *Ion Implantation* | | | | | |
| Friedland (CVD-SiC) [11] | 1200-1400 | $2.4 \times 10^{-9}$ | 3.43 | $1.5 \times 10^{-21}$ | - |
| [b]Friedland (6H-SiC) [8] | 1200-1400 | - | - | [c]$< 1.0 \times 10^{-21}$ | - |
| Friedland(CVD-SiC) [8] | 1200-1400 | $4.3 \times 10^{-12}$ | 2.50 | $4.2 \times 10^{-20}$ | - |
| Maclean (CVD-SiC) [62] | [a]1500 | - | - | - | [a,c]$<5.0 \times 10^{-21}$ |
| Gerczak (CVD-SiC) [14] | [a]1569 | - | - | - | [a]$1.3 \times 10^{-17}$ |
| Leng (CVD-SiC, Unirrad.) [63] | [a]1569 | $1.04 \times 10^{-12}$ | 1.84 | - | [a]$8.8 \times 10^{-18}$ |
| Leng (CVD-SiC, Irrad.) [63] | [a]1569 | - | - | - | [a]$3.6 \times 10^{-17}$ |
| *Diffusion Couple* | | | | | |
| Lopez-Honorato[15] | 950, 1150, [a]1500 | - | - | - | [a]$2.99 \times 10^{-15}$ |
| Lopez-Honorato[12] | 1200 | - | - | [a]$(1.0 - 1600) \times 10^{-18}$ | - |
| Gerczak [14] | 1500 | - | - | - | - |
| **Simulations** | | | | | |
| Shrader(c-SiC) [16] | [d]TST | $6.3 \times 10^{-8}$ | 7.88 | $3.6 \times 10^{-33}$ | $3.9 \times 10^{-29}$ |
| Kahlil ($\sum$3-GB ) [17] | [d]TST | $1.6 \times 10^{-7}$ | 3.95 | $3.5 \times 10^{-17}$ | $3.7 \times 10^{-18}$ |
| Rabone ($\sum$5-GB)[18] | [e]1227 | - | 3.35 | [a]$(0.22 - 10.5) \times 10^{-18}$ | - |
| This work (HEGB) | 1200-1600 | $5.8 \times 10^{-10}$ | 2.79 | $1.6 \times 10^{-18}$ | $3.4 \times 10^{-17}$ |

[a] Diffusion coefficient evaluated at specified temperature
[b] Ag diffusion in single crystal SiC, if marked
[c] Upper limit of diffusion coefficient as no detectable Ag diffusion was observed

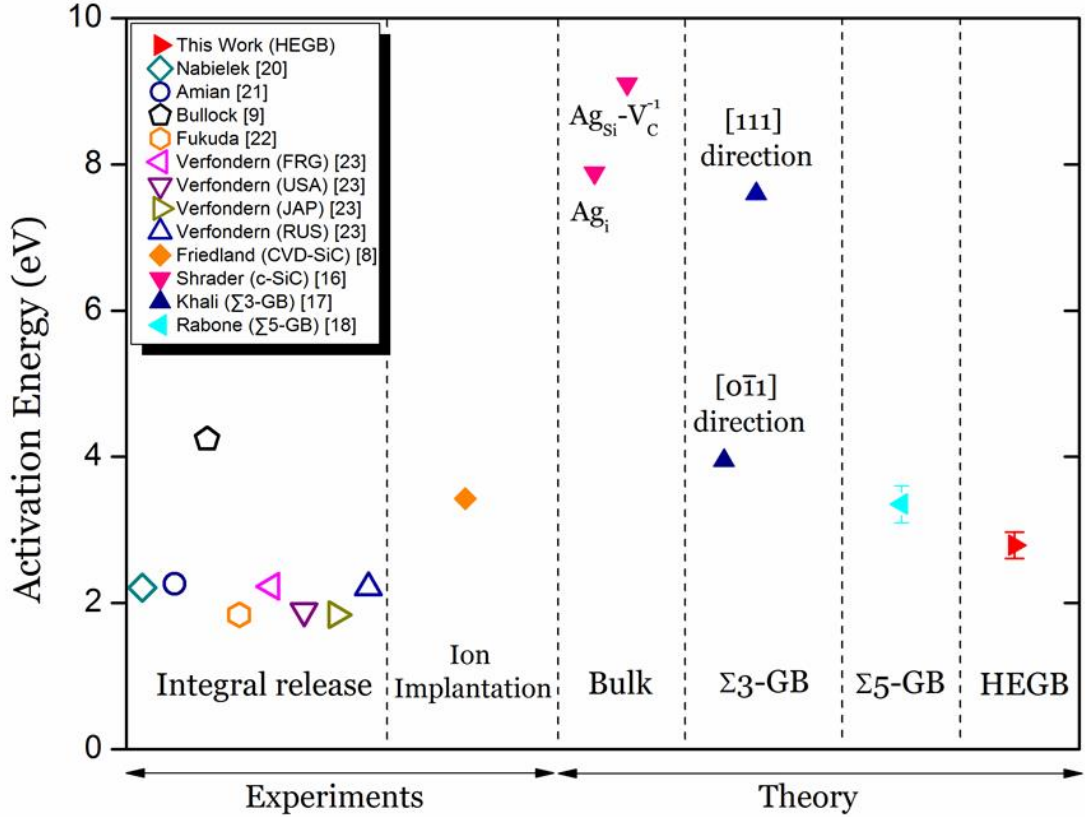

Figure 7. Effective activation energy barrier in this study and literature values [8,9,16-18,20-23]. Note that values predicted in the bulk and ∑3-GB are activation energies for Ag specific species and directions, respectively.

## 4. Discussion

The above results suggest that HEGBs can provide fast diffusion paths. However, it necessary to discuss their contributions to the Ag release in a more general context of other GB related studies and more general aspects of the Ag release process. In particular, the HEGBs are expected to be the dominant Ag diffusion path not only because of their diffusion rate, but also because of the high prevalence and strong Ag segregation tendencies. Here we discuss each of these properties in relation to other GB studies, Ag release studies, and the general Ag release process.

If HEGBs are to provide a fast Ag release path they must enable a connected (percolating) pathway through polycrystalline material. In pc-SiC, ∑3-GBs and random GBs (HEGBs) are the most abundant type of GBs [13,14] and the HEGBs constitute a majority (>40%) of GBs in CVD grown SiC [25]. This value is high enough that it is expected to provide a percolating pathway [17]. It should be noted that of the GB types only HEGBs and ∑3-GBs, with their high fraction of all GBs, can provide a percolation path for Ag diffusion in SiC [17]. Thus, while it is possible that mobility of Ag in other non-∑3 CSL GBs are faster than in HEGB, other paths are less likely to explain accelerated diffusivity given that only HEGB can provide a percolating path for Ag transport. Therefore, it is reasonable to assume that HEGBs may provide the dominant pathways for Ag diffusion in SiC.

If HEGBs are to provide a fast Ag release path, they must also provide stable sites for Ag segregation. A strong segregation of Ag to HEGB is expected based on our calculations, and results to date suggest this segregation is even stronger than for other CSL GBs. The formation energies for substitutional and interstitial Ag (SM, Table S1) are significantly lower than the most stable states of Ag found for bulk (10.5 eV in $Ag_{TC}$). Similarly, the most stable interstitials in the HEGB (which are as low as -0.59 eV) are lower than the interstitial sites found in $\sum$3-GB (5.32 and 2.76 eV (both Ag-V clusters) [17]), and in a $\sum$5-GB (0.88 eV and 1.19 eV.) Ag therefore appears to be notably more stable in the HEGBs than bulk [16] or in other CSLs [17,18], and thus Ag is expected to segregate to these GBs. With a simple non-interacting model for Ag solubility and the Ag interstitial formation energies we predict the Ag solubility limit in the HEGB ($C_{Ag}^{HEGB}$) to be 4.60×10$^{27}$ m$^{-3}$ at 1200 °C ($C_{Ag}$ calculation details are in SM Sec.C.). In sc-SiC and Σ3-GB the solubility limits are calculated as $C_{Ag}^{bulk}$=3.00×10$^{10}$ m$^{-3}$ and $C_{Ag}^{\Sigma 3GB}$ = 3.80×10$^{18}$ m$^{-3}$ at 1200 °C [16,17]. This weak solubility in bulk SiC is consistent with the recent results of Hlatshwayo [64], who observed precipitates of implanted Ag in 6H sc-SiC after thermal exposure at T < 1300 °C, suggesting low bulk Ag solubility. These solubility limit differences are consistent with the Ag strongly segregating to GBs, especially to HEGBs. Furthermore, as show in the SM, our predicted HEGB solubility limit is about 5.1 at. %, which is very close to the values of 2-4 at. % estimated from experimental solubilities [63] assuming they are dominated by HEGBs. This excellent agreement is somewhat fortuitous but further suggests that our Ag formation energetics are accurate.

This work has provided the first guidance on the specific transport properties of HEGBs and suggests that these GBs are a good candidate for being the fastest Ag diffusion paths among different GB types. To further explore whether HEGBs are in fact the dominant mechanism, the $D_{HEGB}$ is compared to previous studies on other mechanisms and rates of Ag diffusion in SiC. A DFT study by Shrader et al. [16] showed extremely low $D_{bulk}$ values (3.9×10$^{-29}$ m$^2$ s$^{-1}$ at 1600°C), suggesting that the GB diffusion is far more important than bulk diffusion, at least in the unirradiated materials. Study on Ag diffusion in a low angle $\sum$3 CSL grain boundary by Khalil et al.[17] has shown a relatively fast diffusion compared to $D_{bulk}$, but still slower than our $D_{HEGB}$ in the range of temperature considered, 1200-1600°C. At 1600 °C, the ratio of $D_{HEGB}$ to $D_{\Sigma 3-GB}$ is close to one, but our result indicates this ratio will increase as the temperature is lowered due to the difference in activation energies. In contrast, Rabone et al.[18] predicted a range of $D_{\Sigma 5-GB}$ to be higher than $D_{HEGB}$ as shown in Figure 6, but their results shows inconsistency with both integral release (open symbols) and ion implantation measurements (non-triangle filled symbols). In fact, there are some potential sources of errors in the approach taken by Rabone et al. [18]. They extracted a migration barrier from a Ag hop in molecular dynamics, which hop was identified from the trajectory and from the change in potential energy. In this approach a single energy barrier for the Ag interstitial hop is used to extract $D$ from the Arrhenius relation. However, it is not clear that this hop is fully representative given that the total energies of the system before and after the hop were found inconsistent. Thus we take $D$ from Rabone et al. to be an upper bound for the diffusion in $\sum$5-GB, as other larger barriers might exist. Although the data is very limited, from the comparison of reported $D$s, Ag in HEGB is expected to diffuse at similar rates or faster than other GB or bulk mechanism in pc-SiC, except perhaps in the ∑5-GB.

In the above arguments, we showed that the HEGBs can provide percolating, highly soluble and fast pathway for Ag transportation. This hypothesis is also supported by experimental studies on Ag diffusion in sc-SiC and a-SiC. In a 6H sc-SiC [7,8], no movement of Ag in SiC was detected at 1300°C [8] and 1500°C [7]. It was postulated that the Ag bulk diffusion is not fast enough to be observed on the experimental time scale, thus the upper bound of $D$ values were estimated from the limited time of experiment. The slow Ag diffusion in sc-SiC is consistent with this work and other simulation studies [16,17]. Particularly encouraging is that the Ag diffusion from amorphized 6H-SiC by Hlatshwayo [64] shows a quantitatively good agreement with our predicted range for a-SiC, although the amorphization was done by ion implantation, which might create a somewhat different amorphous structure than what we model here obtained from molecular dynamics quench. The fair agreement with Ag diffusion in amorphized 6H-SiC experiments not only supports the fast diffusion of Ag in disordered HEGBs but also suggests that the stochastic approach utilized in this work is indeed capable of reproducing realistic Ag transport.

It is useful to compare the $D_{HEGB}$ to net diffusion coefficients in the pc-SiC to assess whether they support HEGB diffusion being the dominant mechanism in pc-SiC. As it can be seen from Figure 6, $D_{HEGB}$ is comparable to, or higher than, the net $D$ values of the recent ion implantation experiments (non-triangle filled symbols) [8,11,63,65] on 3C-SiC. As stated, $D_{\Sigma3-GB}$ and $D_{HEGB}$ are expected to be the preeminent contributors to the net diffusion coefficient because only these GBs have high fractions to form percolating paths. The net $D$ in pc-SiC from existing non-irradiated experimental data showed discrepancies about a factor or ten when compared to $D_{\Sigma3-GB}$ and $D_{HEGB}$, indicating that they are indeed the dominant pathways for Ag diffusion. However, the diffusion couple studies by López-Honorato et al. [12,15] reported $D$ value that is 2 orders of magnitude higher than $D^{HEGB}$, and in fact these values are close to integral release measurements (Figure 6, open symbols) [9,19,20,22-24,66]. In the studies of López-Honorato et al. Ag diffusion was observed with SEM in a heat-treated diffusion couple, where a layer of silver was trapped between CVD-SiC layers. The fast diffusion observed perhaps owes to the higher concentration of Ag, which may support some form of accelerated Ag transport, e.g., through a dissolution type mechanism [67]. Similarly in the Ag/SiC vapor diffusion couple at 1500 °C, Gerczak [14] observed localized dissolution of condensed Ag at SiC surface and Ag penetrated into bulk in Ag-Si corrosion form, not by impurity diffusion kinetics. As the Ag concentration in these experiments is much higher than the expected Ag concentration in the SiC layer of TRISO particles [20,24] the implications of these results for the TRISO particles is not entirely clear.

There is a 2-3 orders of magnitude difference between our predicted $D_{HEGB}$ and the $D$ from integral release measurements in the relevant temperature ranges of 1200-1600 °C. Compared to integral release measurements, other ion implantation experimental observations of Ag diffusion also exhibit a low value, quite consistent with our calculations. These experiments and our calculations are different from the integral release measurements in multiple ways, but perhaps the most obvious difference is that these experiments and our calculations do not involve irradiation. Therefore, we believe that the predicted $D_{HEGB}$ is an approximate upper bound for the effective diffusivity in unirradiated pc-SiC.

As just mentioned, integral release measurements show a clear discrepancy in diffusion rates when compared to most of the other studies, including both experiments (typically ion implantation) and modeling. The discrepancies are likely due to a combination of different factors. However a reasonable explanation for the disagreement of $D$ values between integral release measurements, and ion implantation experiments and models is that radiation affects Ag diffusivity in SiC. In general, radiation enhanced diffusion (RED) has been widely observed in metallic nuclear materials and semiconductors [68-70]. Moreover, a recent study by Leng, *et al.*[63] observed about an one order of magnitude enhancement of Ag diffusion in carbon irradiated pc-SiC, even when the diffusion was measured after the irradiation had stopped. These irradiation effects, if they are occurring, must involve coupling over time as the integral release experiments are generally performed by heating samples that were irradiated previously, so the release is measured long after the irradiation has occurred. It is not clear how the irradiation, especially prior irradiation, might enhance the Ag transport, but below we discuss some possible mechanisms.

Irradiation of SiC not only results elevated point defect concentration but also develops complex defect structures such as interstitial clusters, dislocations, and voids. Irradiation-induced defects may activate different diffusion mechanisms to create RED of Ag, and this might take place in both bulk and GB regions. In irradiated SiC, however, it is likely that GB RED is responsible for the observed fast diffusion of Ag instead of bulk RED, based on the following argument. From earlier discussion we have shown that the Ag solubility in bulk expected to be low, and an extremely strong segregation towards GBs is predicted. Therefore, even if RED occurs in the bulk, the Ag will quickly end up trapped in GBs, which will then provide the dominant diffusion mechanism. Consistent with the hypothesis that the irradiation effects are dominated by RED in GBs, Leng *et al.* [63] showed that the sc-SiC does not show any enhanced Ag transport after irradiation, while irradiated pc-SiC clearly shows RED of Ag at 1400-1569 °C.

The above arguments support the hypothesis that GB RED may bridge the discrepancy of Ag diffusivity between integral release and other reported values. However, there is limited understanding in the possible mechanisms for RED in SiC GBs. For highly disordered GBs, that can be considered approximately amorphous (high-angle GBs), the literature on RED in amorphous materials [71-75] strongly suggests that radiation can change diffusional behaviors. These previously proposed mechanisms includes the collective diffusion via displacement chains [74], enhancement in radiation-induced destabilized regions [71-73], and enhancement by increase of free volume from irradiation [75]. Although there are still many uncertainties about these mechanisms, it is suggested that a RED of Ag through a percolating network of amorphous-like HEGBs may explain the observed radiation effects on Ag diffusion in SiC. In addition to possible RED in disordered GBs, we recently proposed a Ag kickout mechanism [63], that can be active in GBs with crystalline qualities. It was shown in this argument that interstitial clusters formed under irradiation can dissolve during post-irradiation annealing, and then kickout substitutional Ag, leading to its facile interstitial diffusion. Overall it is clear that there are a number of possible mechanisms that could lead to GB RED for Ag, in both more crystalline and more disordered GBs, and further work is needed to explore to what extent which, if any, of these mechanisms are active.

## 5. Conclusion

The *ab initio* calculations are performed to predict the formation and migration energetics of neutral Ag impurities in amorphous SiC (a-SiC) in order model of Ag diffusion in SiC high energy grain boundaries (HEGBs). These models have then been used to better understand Ag release through SiC in TRISO fuels. The mean formation energy for Ag interstitials is 3.5 eV, which is more stable Ag than in crystalline SiC (c-SiC) by 7.0 eV. Thus Ag is predicted to segregate strongly to HEGBs. An interstitial mechanism is predicted to be dominant based on the enhanced stability of Ag interstitials in a-SiC vs. sc-SiC, and low migration barrier compared to other mechanisms.

To model diffusion we first characterized the energy landscape for Ag interstitials in a-SiC. We have searched all Ag interstitial sites in an a-SiC structure and calculated their formation energies ($E_f$), coordination numbers by other Ag interstitial sites ($Z$), and migration barriers between them ($E_{KRA}$). The energy landscape was then combined with a kinetic Monte Carlo simulation to predict the $D_{HEGB}$ at 1200-1600°C. The diffusion of Ag is predicted to have an Arrhenius form with an effective activation energy of 2.79 ± 0.18 eV and a diffusion prefactor of (2.73 ± 1.09) × $10^{-10}$ $m^2s^{-1}$. The diffusion coefficients in this temperature range are comparable with predicted $D_{\Sigma 3\text{-}GB}$ at high temperatures, and are greater at lower temperatures. The consistency between our $D_{HEGB}$ and $D$ measured in the ion implantation experiments on unirradiated polycrystalline SiC implies that Ag is likely to be transported along the grain boundaries pc-SiC. Moreover, $D$ from our amorphized 3C-SiC also showed a good agreement with experimentally measured $D$ from amorphized 6H-SiC, which validates our model. However, combining our results and those from previous studies suggests that diffusion in unirradiated grain boundaries is still too slow to be account for the $D$ measured in integral release experiments. The combined studies to date suggest that we have a fairly consistent understanding of Ag transport through SiC under model conditions, but some other effects are enhancing Ag release in more reactor relevant conditions of integral release measurement. We propose that at least part of the enhanced Ag diffusion is due to radiation enhanced diffusion of Ag in the irradiated SiC, and this is an important area for further investigation in understanding release of Ag from TRISO fuel particles.

## 6. Acknowledgements

This work was supported by the DOE office of Nuclear Energy's Nuclear Energy University Programs under grant No.12-2988. Computations in this work benefitted from the use of the Extreme Science and Engineering Discovery Environment (XSEDE), which is supported by National Science Foundation grant number OCI-1053575.